\documentclass[prd,twocolumn,fleqn,showpacs,showkeys,amsmath]{revtex4}%,preprintnumbers

\usepackage{amsfonts}
\usepackage{epsf}
\usepackage{graphicx}
\usepackage{psfrag}
\usepackage{amssymb}
\usepackage{color}
\usepackage{amsmath}

\newcommand{\rme}{\mathrm{e}}
\newcommand{\rmi}{\mathrm{i}}
\newcommand{\rmd}{\mathrm{d}}

\newcommand{\bfr}{\mathbf{r}}

\newcommand{\cE}{\mathcal{E}}
\renewcommand{\qquad}{\hspace*{25pt}}

\begin{document}

\setcounter{page}{1}%

\title{Bose--Einstein Condensate Dark Matter Model with Three-Particle Interaction
\\ and Two-Phase Structure}

\author{A.M.~Gavrilik\footnote{e-mail: omgavr@bitp.kiev.ua}, M.V.~Khelashvili, A.V.~Nazarenko}
\affiliation{Bogolyubov Institute for Theoretical Physics of NAS of Ukraine, \\ %Nat. Acad. of Sci. of Ukraine
14b, Metrolohichna Str., Kyiv 03143, Ukraine}%

\date{\today}

\begin{abstract}
We explore the consequences of including the repulsive three-particle interaction
in the model of Bose--Einstein condensate dark matter model or fuzzy dark matter.
Such a model based on properly modified Gross--Pitaevskii equation is intended to
describe the distribution of dark matter particles in the highly dense regions,
which correspond to the galaxy core and/or to the overlap of colliding galaxies.
Specifically, we deal with the $\phi^6$-model in terms of the macroscopic wave
function of the condensate, where a locality of interaction is guaranteed by a
large correlation length assumed to hold.  After calculation of main thermodynamical
characteristics, we find strong evidence of the existence of two distinct phases of
dark matter, within its core, separated by the instability region lying between two
differing special values of the pressure acting in the model. Some implications
stemming from the existence of two phases  and the related first-order phase
transition are discussed.

\end{abstract}



\maketitle

\section{Introduction}

%\vspace{-1mm}    {\color{blue}
The notion of dark matter (DM), though being
rather old one~\cite{Zwicky,Bertone}, is a widely accepted concept
nowadays. A support in favor of its existence stems from both
gravitational-lensing observation and the data on galaxy rotation
curves. But, despite a vast amount of theoretical (and experimental)
studies, the ultimate nature of DM remains still unknown.

In the last two decades or more, among a diversity of approaches to
DM, ever-increasing attention is focused on a class of DM models
known as Bose-Einstein condensate (BEC) dark matter (also named ``fuzzy DM'' or FDM,   %equivalently
``quantum wave-function DM'' or $\psi$DM models, scalar-field dark
matter, ultralight dark matter), see~\cite{Sin,Lee,Hu} as well as
the reviews \cite{Matos,Lee17,Hui,Urena,Ferreira,Fan} with numerous
references therein.
 Also, superfluid dark matter models~\cite{Peebles,Berezhiani} are closely
related to this class.   %} %can be included in

\vspace{1mm} Main point/feature of this class of models is that DM
constitutes a BEC subjected to a long range correlation. There are
some variations within this class, depending on (i) the sort and the
values of mass of fuzzy DM constituents, ranging from $~ 10^{-32}$
eV to $~ 10^{-21}$ eV and even higher in some cases,
and on (ii) whether self-interactions are neglected~\cite{Sin,Hu} %%[9]
or the DM particles experience some repulsive self-interaction,
small or not, quartic~\cite{Goodman,Bohmer} or more complex. (The
case with neglected self-interactions is the most restrictive one
and requires the mass of FDM to be $~ 10^{-22}$ eV for the DM core
stability; on the other hand, more freedom is admitted when certain
self-interactions are taken into account.) Among the explored
self-interactions we encounter, besides the most popular quartic
one, also the logotropic~\cite{Chavanis1}, $\cos$- and $\cosh$-type
ones~\cite{Chavanis2,Sahni,Guzman,Matos2}. The studies of the role
of diverse (orders, attractive/repulsive cases of) self-interactions
of ultralight scalar DM particles are of utmost importance.

The FDM class surpasses the cold dark matter models (CDM)
models~\cite{CDM} in the sense that it succeeds to resolve those
problems of CDM that appear at the small (i.e. galactic or
subgalactic) scales. In particular, it was shown that the BEC DM
scenario of DM, exploiting Gross--Pitaevskii (GP) equation related
with quartic term in the  self-interaction scalar potential, jointly
with the Poisson equation, enable to overcome~\cite{Harko2011,Deng}
the well-known cusp/core problem of CDM models.
 Besides, the intrinsic tools of the BEC DM models are efficient enough to
 properly treat the gravitational collapse~\cite{Khlopov}
 issue of the BEC dark matter halos, see e.g.~\cite{Guzman,Harko2}.
 The very important observational realm of distribution of dark matter in galaxies~\cite{Salucci}
 and the galactic rotation curves can as well be described with very good
agreement~\cite{Bohmer,Harko2,BBBS,Kun,Diez2014,Zang2018,Kun2020,Craciun2019,Castellanos2020}.
In a number of works, the BEC DM predictions are tested with the
galaxies kinematic observations using galaxies of diverse
morphologies. It starts from studying of ultracompact dwarf (UCD)
galaxies \cite{Lee2016}, wherein the new scenario of UCD origin was
proposed in  the framework of BEC DM; or the sample of eight
brightest dwarf spheroidal satellites of Milky Way, which was
considered to constrain BEC halo size in the model of BEC DM within
the  Thomas--Fermi approximation~\cite{Diez2014}. The model was
applied as well to the largest galaxies such as Milky Way, that
together with 12 nearby dwarf galaxies from SPARC are used to test
astrophysical effect of BEC DM halo rotation \cite{Zang2018}. It is
worth to mention that SPARC database of disk galaxies of Hubble type
from H0 to Irr (note, the most of galaxies belongs to low surface
galaxies class) is frequently used to compare BEC DM predictions
with the observed galaxies kinematics. The samples of 12 dwarf
galaxies from SPARC \cite{Kun2020} and 139 galaxies of different
types selected from the database according to the data
quality~\cite{Craciun2019} are fitted well by the slowly rotating
BEC DM model. Besides, the BEC DM with super-massive black hole in
halo center  was compared to  NFW density profile by fitting of 20
SPARC galaxies with the desired accuracy of
data~\cite{Castellanos2020}, the both models describe the
observations well, but it is not clear which of two models is
preferred by the data.

In addition, FDM is believed to manage successfully the important
theme of core mass-halo mass relation~\cite{Guzman,Schive1,BBBS}. In
general, the BEC approach to modeling DM, based on (modified
versions of) the Gross--Pitaevskii--Poisson system, helps to
describe the known and reveals possible new features
or phenomena characteristic of DM core and halo.    % rotating galaxy & turbulence - Harko3

It is worth to note that diverse types of Bose-like DM candidates
(satisfying BEC paradigm) have been studied in the literature:
besides the widely explored usual unspecified bosons, there are
rather popular axions (with masses ranging from $~ 10^{-20}$ eV to
$~ 1 $ eV)~\cite{axion,axion2,axion3,axion4,axion5},
        %ultralight axions with a wide range of masses could serve as DM  [15-17].
the Stueckelberg bosons~\cite{Stue}, or even the massive gravitons~\cite{Das,Kun}
(for different bounds on graviton mass, see e.g.~\cite{deRham}).
Moreover, in some recent papers, specific versions of deformed bosons (obeying statistics
differing from the pure Bose one) are explored in the role of DM particles~\cite{Mirza,muBose1,muBose2,Nazar},
and these are admitted thanks to the fact of existence of condensate phase
analogous to usual BEC.
Note that the description of galactic rotation curves obtained within the
particular so-called $\mu$-deformed approach turns out to be quite successful~\cite{muBose2},
even without taking into account the visible (baryonic) component of galaxies.

Recently, the interest is drawn to the role of six-order repulsive
self-interaction term in the scalar potential within the BEC
framework.
Due to sextic term, the situation qualitatively changes~\cite{Luckins}.  %%%%%
In particular, this can lead to nontrivial phase structure of the
core in the central part of the DM halo.  Let us mention the work of
Chavanis~\cite{Chavanis2} which has demonstrated the existence of
two phases  --- one ``dilute'' and the other ``dense'' --- pictured
through the mass-radius relation (note that this was done in the
context of axion stars, with only short remark added that analogous
features can occur in the system of axionic DM).

\vspace{1mm}
  %{\color{blue}
The goal of our paper is to explore main consequences of including
the (repulsive) sextic self-interaction term in the potential of
gravitating BEC dark matter model. The basic aspect of our study
consists in dealing not with the system of Gross--Pitaevskii (GP)
and Poisson equations, but with a single properly modified form of
the GP equation in which  the gravitational potential enters
nonlocally, through the action of inverse Laplacian. That suggests
the extensive usage of numerical methods to study the density
profiles of the BEC dark matter along with their stability
properties. On their base, we are able to obtain the major
thermodynamical functions of BEC DM. Detailed analysis of the
latter, especially of their mutual dependencies, yields the most
valuable information about the appearing phases (stable, metastable,
and unstable regions) and the presence of first-order phase
transition.

\vspace{1mm}
The paper has the following structure: in Sec.~2, main
arguments in favor of the adopted particular choice of the model are
presented, with special attention to the role of sixth order
repulsive self-interaction term in the potential of fuzzy or BEC
dark matter.
  %%%
In Sec.~3, within the appropriately modified version of the
Gross--Pitaevskii equation we treat the sextic self-interaction term
jointly with (the potential of) gravitational interaction employed
in a nonlocal fashion. Due to the latter circumstance, we deal with
single equation instead of the nonlinear Schr\"odinger--Poisson
system. As an important step of the whole analysis, we restrict
ourselves with the Thomas--Fermi approximation (when, due to sending
$\hbar \to 0$, kinetic term is eliminated).
  %%%
Then, the complete form of the model is explored in Sec.~4.
Namely, with inclusion of quantum processes, the  calculation of the
relevant thermodynamical functions is performed.
On this base, in Sec.~5 we establish the existence of, and analyze in detail,
the two distinct phases that are realized in the dark matter core.
The final section is devoted to discussion of implications, concluding
remarks and outlook.

%% %% %% %% %% %% %% %% %% %% %% %% %% %% %% %% %% %% %% %% %% %% %% %% %% %% %% %%
\section{Constructing a Model}

In this Section, we present the criteria that allow us to formulate
a model of dark matter (together with the conditions of its
applicability), which we then further study. Our arguments stem from
a comparison of empirical data for galaxies and the properties of
some theoretical models of dark matter formed by Bose particles (and
fields).

\subsection{The Scaling Criterion for Interactions}

As a kind of the criteria that justifies the ability of a
theoretical model to describe observables, we use scaling
reasonings. By implementing the scaling in practice, we intend to
choose a more realistic model for our study.

Considering a sample of observed galaxies in a wide mass range, it
allows us to relate core size $r_0$ and mass density $\rho_0$ in the
center. Such a relation is found to be close to $\rho_0\propto
r_0^{-\beta}$ with $\beta\approx 1$. In fact, $\beta=1.3$ for the
studied galaxies sample. A close result is obtained in
\cite{Chen2017} from analysis of {\it dwarf spheroidal} galaxies,
which gives a similar correlation between halo radius and mass
density, that is, $\rho_0\propto r_0^{-1.2}$.

Let us first review the scaling relations of {\it fuzzy dark matter}
(FDM), where DM particles are described by complex-valued wave
function $\psi(r)\,\rme^{-\rmi\gamma t/\hbar}$ and individual mass
$m$. Accordingly to FDM, stationary Schr\"odinger--Poisson equations
are
\begin{equation}
\gamma \,\psi = -\frac{\hbar^2}{2m}\Delta\psi + m\,V_{\mathrm{gr}}\,\psi,\quad
\Delta V_{\mathrm{gr}}(r)=4\pi G \rho(r),
\end{equation}
where $\Delta$ is a Laplace operator; $\rho=m|\psi|^2$.

The system possesses exact scaling symmetry with parameter
$\lambda$:
\begin{equation}
r \to \lambda^{-1} r, \quad \rho\to\lambda^4 \rho\,,
\quad
\psi\to\lambda^2 \psi, \quad \gamma\to\lambda^2\gamma\,.
\label{sclaing-symmetry}
\end{equation}

It leads to the scaling relation $\rho_0 \propto r_0^{-4}$ between
DM halo radius $r_0$ and central density $\rho_0$. As it is shown in
\cite{Deng}, this relation strictly contradicts the observations.
Indeed, $\beta=4$ is far from $\beta=1.3$ already noted above.
Although, a realistic model should be applicable to the DM
description, as is expected, for a large sample of galaxies, but not
only exceptions.

An important conclusion results from the FDM model: In order to reach $\beta\simeq1$
we need to take self-interaction among DM particles into account.

Including repulsive or attractive $\psi^4$ self-interaction into the
FDM model, the scaling relation takes the form with $\beta=2$ for a
large range of disk size, that is close to the value derived in
\cite{Deng}. However, the solutions found are unstable and thus do
not resolve the issue. Nevertheless, there is a possibility to
obtain stable solutions in the required range of parameters,
combining repulsive and attractive potentials in some way to prevent
a halo collapse or outward flow of particles.

Continuing the analysis, let us now test $\psi^6$-model in details. Clearly, the FDM equation
with repulsive $\psi^6$ term does not obey the scaling symmetry \eqref{sclaing-symmetry}
any more as well as any other scaling relations. Thus, it becomes more complicated to find
relation between $r_0$ and $\rho_0$.

We are forced to apply a hint, indicating that solution of problem
can be found within the framework of $\psi^6$-model. Following
\cite{axion5}, we estimate the contribution of several terms to a
total energy of halo using an exponential ansatz. This trick has
already been used in the literature to study a stability of scalar
field DM halos.

Thus, let the total energy $E$ (in infinite volume, with its element
$\rmd V(r)=4\pi r^2\,\rmd r$) and its density $\cE$ of spherically
symmetric DM halo be
\begin{equation}
E=\int_0^\infty\mathcal{E}(r)\,\rmd V(r),\quad
\cE=\cE_{q}+\cE_{\mathrm{int}}+\cE_{\mathrm{gr}},
\label{energy-functional}
\end{equation}
where
\begin{equation}\label{Eterms}
\cE_{q}=\frac{\hbar^2}{2m}\left|\frac{\rmd\psi}{\rmd r}\right|^2,\quad
\cE_{\mathrm{int}}=\frac{U}{3}|\psi|^6,\quad
\cE_{\mathrm{gr}}=\frac{1}{2}\rho\, V_{\mathrm{gr}}.
\end{equation}
The wave function $\psi$ is normalized so as to give the total
number of particles $N$:
\begin{equation}\label{norm}
N=\int_0^\infty |\psi(r)|^2\,\rmd V(r).
\end{equation}

Further, we adopt the exponential ansatz for $\psi$:
\begin{equation}
\Phi(r) = \sqrt{\frac{N}{\pi r_0^3}}\, \rme^{-\frac{r}{r_0}},
\label{ansatz}
\end{equation}
which satisfies (\ref{norm}).

This is a simple ansatz to simulate field behavior in the inner region when an exact solution
is unknown. It is valid in the range of $r \lesssim r_s$ for some finite $r_s$, but breaks down
in the outer region.

Substituting (\ref{ansatz}) instead of $\psi$ into \eqref{energy-functional}, one gets
\begin{eqnarray}
&&E=a\,\frac{N}{r_0^2}+b\,\frac{N^3}{r_0^6}-c\,\frac{N^2}{r_0},\\
&&a=\frac{\hbar^2}{2m},\quad b=\frac{U}{81\pi^2},\quad c=\frac{5G m^2}{16}.
\end{eqnarray}
To calculate the action of operator $\Delta$ and its inverse on function $f(r)$
in the spherically symmetric case (what is needed for finding $V_{\mathrm{gr}}$),
it is enough to use their radial parts defined as
\begin{eqnarray}
&&\Delta_rf(r)=\partial^2_rf(r)+\frac{2}{r}\,\partial_rf(r),\label{Dlt}\\
&&\Delta^{-1}_rf(r)=-\frac{1}{r}\int_0^rf(s)\,s^2\,\rmd s-\int_r^{R}f(s)\,s\,\rmd s.
\label{InvDelta}
\end{eqnarray}
where $R$ is radius of the ball, where the matter is located. We put $R\to\infty$ here.

Energy $E$ can be presented in the form:
\begin{eqnarray}
&&E=b\,\frac{N}{r^6_0}(N-N_-)(N-N_+),\\
&&N_\pm=\frac{c\,r^5_0}{2b}\left[1\pm\sqrt{1-\frac{4ab}{c^2r^6_0}}\right]>0.
\end{eqnarray}
In the bound state, it should be $-\infty<E<0$ what leads to condition $N_-<N<N_+$.
Evaluating $N$ as geometric mean $\sqrt{N_-N_+}$, we obtain that $N \propto r_0^2$.

Now we can estimate the central density:
\begin{equation}
\rho_0 \propto \frac{N}{r_0^3} \propto r_0^{-1}.
\end{equation}

This dependence with $\beta=1$ is obtained through a series of rough
approximations and does not claim to be a final answer. However, the
estimation obtained here is encouraging and stimulates us to study
the model with potential $\psi^6$.

%% %% %% %% %% %% %% %% %% %% %% %% %% %% %% %% %% %% %% %% %% %% %% %% %% %% %% %%
\subsection{``Core+Tail'' Profile}

Since the $\psi^6$-model actually takes into account the
three-particle interaction that is relevant under certain conditions
in a small region of space (like a core), finding the distribution
of dark matter in a large region of space requires an extension of
the model. To extend this, we borrow ideas already applied to other
models.

The large-scale structure simulation within the $\psi$DM model shows
that there are structures (like filaments and voids) which are
similar to the CDM outcomes. At the same time, DM distribution on a
smaller, galactic scales differs due to the wave nature of $\psi$DM.
Its prominent feature is formation of the coherent standing waves of
DM that form a flat core in the center of galactic DM halo. Such
cores are described very well by the soliton-like solution of
Schr\"odinger--Poisson equation also known as {\it boson star
solution}. These cores are also surrounded by an envelope of
incoherent phase, that mimics Navarro--Frenk--White (NFW) profile
\cite{Schive2014,Hui}.

Therefore, $\psi$DM profile is often represented as
\begin{equation}\label{CTpr}
\rho(r)=\rho_{core}(r)\,\theta(r_a-r)+\rho_{NFW}(r)\,\theta(r-r_a) \,,
\end{equation}
where $\theta(r)$ is a Heaviside theta-function. At the point $r=r_a$,
we require $\rho_{core}(r_a)=\rho_{NFW}(r_a)$. Value of $r_a$ is selected,
for instance, to relate the mass of central soliton to the mass of whole DM halo as
\begin{equation}
M_{s} \propto M_{halo}^{1/3}\,.
\end{equation}
This relation occurs in \cite{Schive2014,Schive1} and reflects also
equality of the energy per unit mass for the central core and the
NFW envelope \cite{BBBS}.

\begin{figure}[htbp]
\vspace*{-2mm}
\includegraphics[width=7.8cm,angle=0]{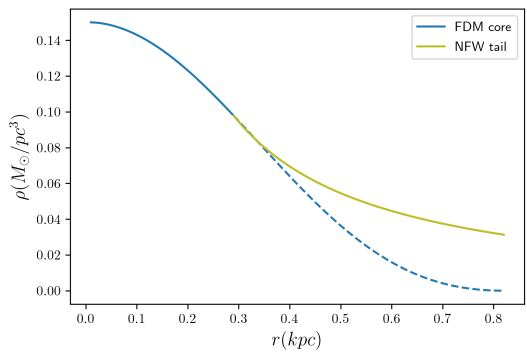}
%\vspace*{-2mm}
\caption{\label{FDM_NFW} The numerical FDM core solution with the NFW outer part.}
\end{figure}

A particular ``extension'' (\ref{CTpr}) of the FDM model, which is described by
\begin{eqnarray}
&&-\Delta_r\psi(r)+\varkappa U_{\mathrm{gr}}(r)\,\psi(r)=\tilde\gamma\,\psi(r),\\
&&U_{\mathrm{gr}}(r)=-\frac{1}{r}\int_0^r\psi^2(s)\,s^2\,\rmd s-\int_r^{\infty}\psi^2(s)\,s\,\rmd s,
\end{eqnarray}
is sketched in Fig.~\ref{FDM_NFW} for central density $\rho_0=0.15\ M_{\odot}\,\mathrm{pc}^{-3}$
and particle mass $m=1.6\cdot10^{-23}\ \mathrm{eV}\,c^{-2}$;
\begin{equation}
\varkappa=\frac{8\pi G}{\hbar^2}\,\rho_0m^2r^4_*\simeq1.106;\quad r_*=1\ \mathrm{kpc}.
\end{equation}

We would like to prove that a form (\ref{CTpr}) of DM profile, consisting of both soliton-like solution to
the field equations in the inner region and NFW tail in the outer region of halo, remains correct for the
self-interacting field. It is justified by evaluating the contribution of each term to the total energy.
The same can be done by considering a ratio of these terms in the Schr\"odinger--Poisson equations.

Let us consider the decay rate $g(r)$ for the self-interaction $\cE_{\mathrm{int}}$
with respect to the gravitational term $\cE_{\mathrm{gr}}$, which are given by (\ref{Eterms}).
Thus, we introduce
\begin{equation}
g(r)=\frac{q(r)}{q(0)},\qquad q(r)=\frac{\cE_{\mathrm{int}}(r)}{\cE_{\mathrm{gr}}(r)}.
\end{equation}

Evaluating $g(r)$, we use (\ref{ansatz}) again and obtain
\begin{equation}
g(r)=\frac{V_\mathrm{gr}(0)}{V_\mathrm{gr}(r)}\,\rme^{-4\frac{r}{r_0}}
\simeq\frac{r}{r_0}\,\rme^{-4\frac{r}{r_0}},\quad r>r_0.
\end{equation}

On the other hand, there is the inner region $r<r_s$ with
$g(r)\sim1$, where $r_s$ is some characteristic radius. This region
should contain rather $1/4$ of total halo mass like in a
noninteracting field model.

Since $g(1.2\,r_0)\approx 0.01$ and $g(1.9\,r_0)\approx 0.001$,
self-interaction can be neglected with a high accuracy in the outer region of halo.
At the same time, the NFW approximation remains valid for the outer part of halo even
in the case of a self-interacting field.

%% %% %% %% %% %% %% %% %% %% %% %% %% %% %% %% %% %% %% %% %% %% %% %% %% %% %% %%
\section{The case of Tomas--Fermi Approximation}

Although the distribution (\ref{CTpr}) is promising from the point
of view of observations, a $\psi^6$-model and its properties are of
independent interest. In fact, we intend to study the boson
subsystem formed by gravity and three-particle interaction, which
can prevail at a high particle density in a small region of space
like a galactic core.

Let us start from a macroscopic model of gravitating Bose--Einstein
condensate with three-particle interaction, limiting ourselves by
the spherically symmetric case and by the absence of hydrodynamic
flows.

Introducing a constant chemical potential $\tilde\mu$, we will describe the condensate by real
function $\psi(r)$ of radial variable $r=|\bfr|$. Thus, a starting point of our study is
the energy functional in a ball $B=\{\bfr\in\mathbb{R}^3|\, |\bfr |\leq R\}$:
\begin{eqnarray}
&&\Gamma=4\pi\int_0^R\left[\frac{\hbar^2}{2m}(\partial_r\psi(r))^2
+m\psi^2(r)V_{\mathrm{ext}}(r)\right.\nonumber\\
&&\hspace*{6mm}
\left.+\frac{U}{3}\psi^6(r)-\tilde\mu\psi(r)^2\right]\,r^2\,\rmd r;\label{G1}\\
&&\Delta_r V_{\mathrm{ext}}(r)=4\pi Gm|\psi(r)|^2,
\end{eqnarray}
where $\Delta_r$ is the radial part of Laplace operator (\ref{Dlt}).

For the sake of simplicity, let us introduce dimensionless variables:
\begin{eqnarray}
&&\psi(r)=\sqrt{\varrho_0}\,\chi(\xi),\quad r=r_0\,\xi,\nonumber\\
&&A=4\pi\frac{Gm^3\varrho_0 r_0^4}{\hbar^2},\quad
B=U\frac{\varrho_0^2 r_0^2 m}{\hbar^2},\quad u=\tilde\mu\,\frac{m r_0^2}{\hbar^2},
\label{param}
\end{eqnarray}
where $\chi(\xi)$ is a real dimensionless field; $\varrho_0$ and $r_0$ characterize
{\it typical measures} of the central particle density and the system size,
respectively.

Thus, we arrive at
\begin{eqnarray}
&&\hspace*{-3mm}
\frac{\Gamma}{\Gamma_0}=\int_0^{\xi_B}\left[\frac{1}{2}(\partial_\xi\chi)^2-u\chi^2
+A\chi^2\varphi+\frac{B}{3}\chi^6\right]\xi^2\,\rmd\xi,
\nonumber\\
&&\hspace*{-3mm}
\Gamma_0=\frac{4\pi\hbar^2r_0\varrho_0}{m},\qquad
\Delta_{\xi}\varphi(\xi)=\chi^2(\xi),\label{G2}
\end{eqnarray}
where $R=r_0\xi_B$; $\Delta_{\xi}$ and $\Delta^{-1}_{\xi}$ are given
by (\ref{Dlt}), (\ref{InvDelta}) in terms of $\xi$ replacing $r$; \
also, $\chi(\xi)=\psi(r)/\sqrt{\varrho_0}$, see (\ref{param}).

It is useful to evaluate immediately the range of model parameters. Turning to
the known data for galactic cores, we assume that the central mass density
$\rho_0=m\varrho_0$ is of the order of magnitude $10^{-20}\ \mathrm{kg}\,\mathrm{m}^{-3}$
and the light-boson mass $m$ is of the order of $10^{-22}\ \mathrm{eV}\,c^{-2}$. Further, we
use a definition of parameter $A$ to determine the characteristic radius $r_0$,
defining a total radius $R=r_0\xi_B$. One obtains
\begin{eqnarray}
r_0&\simeq& 0.824\ \mathrm{kpc}\,\left[\frac{A}{10}\right]^{1/4}
\,\left[\frac{mc^2}{10^{-22}\ \mathrm{eV}}\right]^{-1/2}
\nonumber\\
&&
\times\left[\frac{\rho_0}{10^{-20}\ \mathrm{kg}\,\mathrm{m}^{-3}}\right]^{-1/4}.
\label{rA}
\end{eqnarray}
Since the realistic values of $r_0$ are smaller than 1~kpc, we estimate
the measure of gravitational interaction as $A\sim10$. Note that (\ref{rA})
cannot be regarded as relation between $r_0$ and $\rho_0$ discussed in the
previous Section.

While the gravity looks like a cumulative effect of a whole system, (thermo)dynamics
of internal processes is strongly determined by repulsive interaction among bosons,
represented by parameter $B>A$.

The characteristic energy density is $\varepsilon_0=\hbar^2\varrho_0/(m r^2_0)$.
Combining this formula with (\ref{rA}), one gets
\begin{eqnarray}
\varepsilon_0&\simeq& 33.82\ \mathrm{eV}\,\mathrm{cm}^{-3}\,\left[\frac{A}{10}\right]^{-1/2}
\,\left[\frac{mc^2}{10^{-22}\ \mathrm{eV}}\right]^{-1}
\nonumber\\
&&
\times\left[\frac{\rho_0}{10^{-20}\ \mathrm{kg}\,\mathrm{m}^{-3}}\right]^{3/2}.
\label{eA}
\end{eqnarray}
In the pressure units, $33.82\ \mathrm{eV}\,\mathrm{cm}^{-3}\simeq 5.42\cdot10^{-12}\ \mathrm{Pa}$.

A detailed analysis of the model based on $\Gamma$ will be performed
in the next Section.
 However, before exploring the model in most
general situation and in order to get an idea of the basic
properties of the model, we first turn to the Thomas--Fermi
approximation (at $\hbar\to0$):
\begin{equation}\label{TF}
\frac{\Gamma_\mathrm{TF}}{\Gamma_0}=\int_0^{\xi_B}\left[-u\,\eta
+A\,\eta\varphi+\frac{B}{3}\,\eta^3\right]\xi^2\,\rmd\xi,
\end{equation}
where $\eta(\xi)=\chi^2(\xi)$ determines a local particle density.

The particle distribution $\eta(\xi)$ is found by extremizing functional $\Gamma_\mathrm{TF}$,
$\delta\Gamma_\mathrm{TF}/\delta\eta(\xi)=0$, that gives us an integral equation:
\begin{equation}\label{eq1TF}
B\eta^2(\xi)+A\varphi(\xi)-u=0,\qquad
\varphi(\xi)=\Delta^{-1}_{\xi}\eta(\xi).
\end{equation}

We are interested in a solution which satisfies the following conditions:
\begin{equation}\label{eqTF}
\eta(0)=\eta_0,\qquad \eta^\prime(0)=0,\qquad \eta(\xi_B)=0.
\end{equation}

A chemical potential $\mu(\xi)$ of the system in gravitational field \cite{LL}
is such that
\begin{equation}\label{mTF}
\mu(\xi)+A\varphi(\xi)=u,
\end{equation}
which is needed for further purposes. Taking (\ref{eq1TF}) into account, $\mu(\xi)$
determines a particle density $\eta(\xi)$ as
\begin{equation}
\mu(\xi)=B\eta^2(\xi),\qquad
\mu(\xi_B)=0.
\end{equation}

Combining, $\mu(\xi)$ satisfies an equation:
\begin{equation}\label{muTF}
\Delta_\xi\mu=-\frac{A}{\sqrt{B}}\,\mu^{1/2},\quad
\mu(0)=B\eta^2_0,\quad \mu^\prime(0)=0.
\end{equation}

To compute $\eta$ (and $\mu$), we first transform $\varphi(\xi)$ as
\begin{eqnarray}
\varphi(\xi)&=&-\frac{1}{\xi}\int_0^\xi \eta(s)\,s^2\,\rmd s
-\int_\xi^{\xi_B} \eta(s)\,s\,\rmd s\nonumber\\
&=&-\upsilon(\xi_B)+\frac{1}{\xi}\int_0^\xi \upsilon(s)\,\rmd s,
\label{trans1}
\end{eqnarray}
where an auxiliary field $\upsilon(\xi)$ is defined by
\begin{equation}\label{ups}
\partial_\xi\upsilon(\xi)=\xi\,\eta(\xi),\quad \upsilon(0)=0;\quad
\upsilon(\xi)=\int_0^\xi\eta(s)\,s\,\rmd s.
\end{equation}

Thus, Eq.~(\ref{muTF}) is solved at $\mu(0)=\nu$, where new parameter
\begin{equation}
\nu=A\,\upsilon(\xi_B)+u>0
\end{equation}
absorbs unknown $\upsilon(\xi_B)$ due to arbitrariness of $u<0$.

Transformation (\ref{trans1}) allows us to rewrite
$\Gamma_\mathrm{TF}$ in alternative form:
\begin{eqnarray}
&&\frac{\Gamma_\mathrm{TF}}{\Gamma_0}=\int_0^{\xi_B}\left[-u\,\eta(\xi)
+\frac{B}{3}\,\eta^3(\xi)\right]\,\xi^2\,\rmd\xi\nonumber\\
&&\hspace{11mm}
-\frac{A}{2}\int_0^{\xi_B} [\upsilon(\xi_B)-\upsilon(\xi)]^2\,\rmd\xi.
\label{TF1}
\end{eqnarray}
Here, the contributions of repulsive and attractive interactions are easily recognized.

\begin{figure}[htbp]
\includegraphics[width=7.7cm,angle=0]{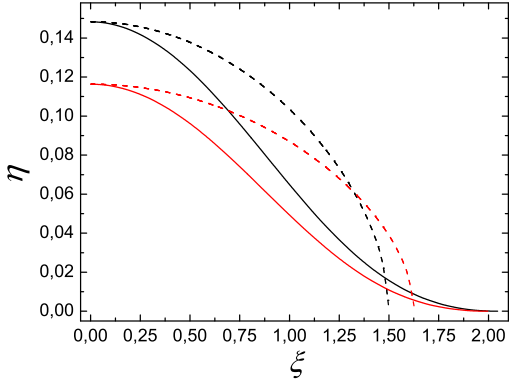}
\vspace*{-2mm} \caption{\label{dens} Spatial development of particle
density $\eta(\xi)=\chi^2(\xi)$ in the models with both quantum
fluctuations (solid) and in TF approximation (dashed) at $A=10$ and
the same initial condition. Black and red curves correspond to
$B=20$ and $B=30$, respectively. For solid curves see the text after
Eq.(60) below.}
\end{figure}

To obtain (\ref{eq1TF}) by varying (\ref{TF1}) with respect to $\eta$, we should take
the relations (\ref{ups}) into account. Thus, using our notations, (\ref{eq1TF}) can be
reduced to the set of equations:
\begin{eqnarray}
&& \partial_\xi\upsilon(\xi)=\xi\,\eta(\xi),\qquad \upsilon(0)=0, \label{TFeq1}\\
&& \eta(\xi)=\sqrt{\eta^2_0-\frac{k}{\xi}\int_0^\xi \upsilon(s)\,\rmd s},\qquad \eta_0=\sqrt{\frac{\nu}{B}},
\label{TFeq2}
\end{eqnarray}
where parameters $\eta_0$ and $k=A/B$ should be given. Equation (\ref{TFeq2})
indicates that function $\eta(\xi)$ vanishes at some $\xi=\xi_B$, depended
on the parameters used (see Fig.~\ref{dens}, dashed curves).

In a consistent way, we get at the boundary $\xi=\xi_B$:
\begin{equation}\label{unu}
\frac{A}{\xi_B}\int_0^{\xi_B} \upsilon(\xi)\,\rmd\xi=\nu,\quad
u=-\frac{A}{\xi_B}\int_0^{\xi_B} \eta(\xi)\,\xi^2\,\rmd\xi.
\end{equation}
Therefore, a negative chemical potential $u$ coincides with the (dimensionless)
gravitational potential of a whole system at the boundary.

We solve numerically the set of equations (\ref{TFeq2}), using the Euler's
method, to find fields $\eta(\xi)$ and $\upsilon(\xi)$. It is because this
set is not integrable analytically.

Particularly, at $A=B=\nu$, what can be achieved by rescaling, the
problem coincides with the Lane--Emden equation with polytopic index $p=1/2$ for
$\theta(\xi)=\eta^2(\xi)$ (see (\ref{muTF})):
\begin{equation}
\Delta_\xi\theta(\xi)=-\theta^{p}(\xi), \qquad \theta(0)=1.
\end{equation}
Note that it has $\xi^{exact}_B\simeq2.75269805$ \cite{Horedt}.

In general case, we are interested in looking for macroscopic
characteristics of the system like an extreme $\Gamma^{ex}_\mathrm{TF}$,
obtained from $\Gamma_\mathrm{TF}$ by inserting the solution $\eta(\xi)$.

Technically, varying $A$, $B$ and $\nu$, functions $\eta(\xi)$ and $\upsilon(\xi)$ (together
with the values of $\xi_B$ and $\upsilon(\xi_B)$) can be found (numerically) in accordance with
(\ref{TFeq1}) and (\ref{TFeq2}). Using the relation $u=\nu-A\,\upsilon(\xi_B)$ and (\ref{TF1}),
we are able to compute $\Gamma^{ex}_\mathrm{TF}$ and the others.

Finding macroscopic characteristics, we appeal to the thermodynamic relations at $T=0$:
\begin{eqnarray}
\rmd p(\xi)&=&\eta(\xi)\,\rmd\mu(\xi),\hspace*{17.2mm} p(\xi_B)=0,\label{GDr}\\
\varepsilon(\xi)&=&\eta(\xi)\,\mu(\xi)-p(\xi),\qquad \varepsilon(\xi_B)=0,\label{Eu}
\end{eqnarray}
where functions $p(\xi)$ and $\varepsilon(\xi)$ determine the (dimensionless)
mean pressure $P$ and the internal energy $E$:
\begin{equation}
P=\frac{3}{\xi^3_B}\int_0^{\xi_B}p(\xi)\,\xi^2\,\rmd\xi,\qquad
E=\int_0^{\xi_B}\varepsilon(\xi)\,\xi^2\,\rmd\xi.
\end{equation}
Hereafter, $\xi^3_B/3$ represents the volume of the system.

Therefore, we need to integrate first the Gibbs--Duhem relation
(\ref{GDr}) and to substitute $p(\xi)$ into the Euler relation
(\ref{Eu}) in order to find $\varepsilon(\xi)$. Thus, one obtains
the explicit expressions:
\begin{equation}\label{EPtf}
p(\xi)=\frac{2}{3}B\eta^3(\xi),\qquad
\varepsilon(\xi)=\frac{1}{3}B\eta^3(\xi),
\end{equation}
which give us the equation of state by inserting the solution $\eta(\xi)$ of (\ref{TFeq1})--(\ref{TFeq2}).

It is instructive to express the {\it gravitational energy} $E_{\mathrm{gr}}$,
\begin{equation}
E_{\mathrm{gr}}=\frac{A}{2}\int_0^{\xi_B}\eta(\xi)\varphi(\xi)\xi^2\rmd\xi,
\end{equation}
the {\it internal energy} $E$ and the {\it total energy}
$E_{\mathrm{tot}}=E+E_{\mathrm{gr}}$ in terms of
gravitational potential $u$ at the boundary:
\begin{equation}
u\mathcal{N}=-A\,\frac{\mathcal{N}^2}{\xi_B},\qquad
\mathcal{N}=\int_0^{\xi_B} \eta(\xi)\,\xi^2\,\rmd\xi.
\end{equation}

Using the auxiliary notations and formulas:
\begin{eqnarray}
&&n(\xi)=\int_0^{\xi}\eta(s)\,s^2\,\rmd s,\quad \mathcal{N}=n(\xi_B),\\
&&\varphi(\xi)=-\frac{\mathcal{N}}{\xi_B}-\int_{\xi}^{\xi_B}\frac{n(s)}{s^2}\,\rmd s,\\
&&\partial_\xi\left(\frac{n^2(\xi)}{\xi}\right)+\frac{n^2(\xi)}{\xi^2}=2\frac{n(\xi)\,\partial_\xi n(\xi)}{\xi},\\
&&\frac{1}{\eta(\xi)}\partial_{\xi}p(\xi)=\partial_\xi\mu(\xi)=-A\frac{n(\xi)}{\xi^2},\label{Pev}
\end{eqnarray}
one obtains
\begin{eqnarray}
&&\int_0^{\xi_B}p(\xi)\xi^2\rmd\xi=\frac{2A}{9}\frac{\mathcal{N}^2}{\xi_B},
\quad E_{\mathrm{gr}}=-\frac{2A}{3}\frac{\mathcal{N}^2}{\xi_B},\\
&&E=\frac{A}{9}\frac{\mathcal{N}^2}{\xi_B},\quad
E_{\mathrm{tot}}=-\frac{5A}{9}\frac{\mathcal{N}^2}{\xi_B},
\end{eqnarray}
which agree with the general results \cite{LL} for gravitating systems with polytropic
equation of state at $T=0$.

%% %% %% %% %% %% %% %% %% %% %% %% %% %% %% %% %% %% %% %% %% %% %% %% %% %% %% %%
\section{The Processes Driven by Quantum Fluctuations}

Let us return to the model with the included quantum kinematics, which is initially described
by functional (\ref{G2}). The field equations read
\begin{equation}\label{teq1}
\frac{1}{2}\Delta_\xi\chi+u\chi-A\chi\varphi-B\chi^5=0,\qquad
\Delta_{\xi}\varphi=\chi^2.
\end{equation}

We combine the model equations in the spirit of the previous Section
by introducing field $\upsilon(\xi)$. It gives us
\begin{eqnarray}
&& 2\frac{\Gamma}{\Gamma_0}=\int_0^{\xi_B}\left[(\partial_\xi\chi)^2-u_*\,\chi^2(\xi)
+\frac{B_*}{3}\,\chi^6(\xi)\right]\,\xi^2\,\rmd\xi\nonumber\\
&&\hspace*{10mm}-\frac{A_*}{2}\int_0^{\xi_B} [\upsilon(\xi_B)-\upsilon(\xi)]^2\,\rmd\xi,\label{g1}\\
&& \Delta_\xi\chi+\nu\chi-\chi \frac{A_*}{\xi}\int_0^\xi\upsilon(s)\,\rmd s-B_*\chi^5=0,\label{g2}\\
&& \partial_\xi\upsilon(\xi)=\xi\,\chi^2(\xi),\qquad \upsilon(0)=0,\label{g3}\\
&& \nu=A_*\upsilon(\xi_B)+u_*,\label{g4}
\end{eqnarray}
where $A_*=2A$, $B_*=2B$ and $\nu$ (instead of $u_*=2u$) are
arbitrary positive parameters. The system boundary $\xi_B$ is
defined from condition $\chi(\xi_B)=0$ and is the {\it first zero}
of oscillating function $\chi(\xi)$. Of course, the values of
$\xi_B$ would differ from those in the TF approximation at the same
$A$ and $B$ that can be seen in Fig.~\ref{dens}.

Aiming to obtain a solution with a finite initial value $\chi_0=\chi(0)<\infty$, it is naturally
to require $\chi^\prime(0)=0$. In order to obtain a decaying solution (finite for admissible $\xi$),
we formulate the following conditions which fix $\chi_0$.

Expanding $\chi(\xi)=\chi_0+C_2\xi^2+\dots$ at $\xi\to0$ and substituting it in (\ref{g2}), (\ref{g3}),
the set of algebraic equations arises:
\begin{eqnarray}
&&6C_2+\nu \chi_0-B_* \chi^5_0=0,\nonumber\\
&&\nu C_2-\frac{A_*}{6}\,\chi^3_0-5B_* \chi^4_0 C_2=0.
\label{ICs}
\end{eqnarray}

Combining, the initial value $\chi_0$ should satisfy the equation $S(A_*,B_*,\nu,\chi_0)=0$, where
\begin{equation}
S(A_*,B_*,\nu,z)=A_* z^2-(5B_*z^4-\nu)(\nu-B_*z^4),
\end{equation}
together with condition $2C_2=\chi^{\prime\prime}(0)\leq0$. These constraints limit
$\chi_0$ as $(\nu/5B_*)^{1/4}<\chi_0<(\nu/B_*)^{1/4}$.

In practice, three regimes (for given $A_*$, $B_*$ and $\nu$) are observed: 1) no solution;
2) single solution; 3) pair of (positive) solutions. Usually, for fixed $A_*$ and $B_*$,
but increasing $\nu$, the indicated sequence of all three modes occurs.

The absence of a solution of (\ref{ICs}) in step 1) leads to the only possible solution
to the autonomous equation (\ref{g2}): $\chi(\xi)=0$. At the stage 2), we obtain a
{\it minimal admissible value} $\nu_{\mathrm{min}}$. Starting from this threshold,
the system begins to evolve. At the stage 3), one of the values of $\chi_0$, which
corresponds to a minimal of these, should be chosen (because the other leads to
divergent $\chi(\xi)$).

Actually, all quantities of the model, computed at fixed $A$ and $B$, are supposed
to be functions of free parameter $\nu$. Thus, dependence, say, of $a$ on $b$ should be
treated in parametric form: $a(b)=\{(b(\nu),a(\nu))|\nu\geq\nu_{\mathrm{min}}\}$.

\begin{figure}[htbp]
\includegraphics[width=7.0cm,angle=0]{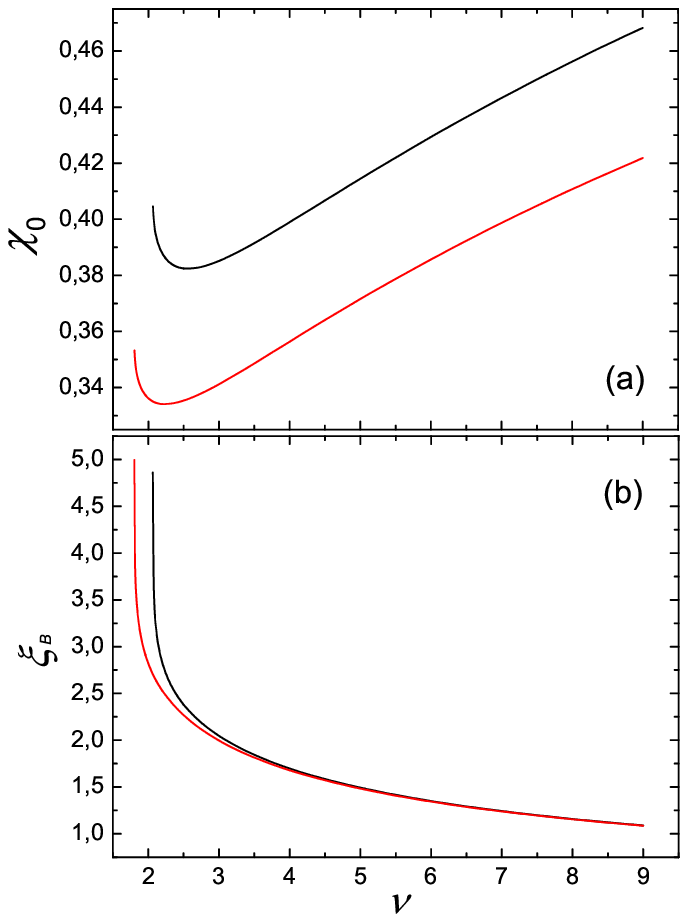}
\vspace*{-2mm}
\caption{\label{chi0} Characteristics of $\chi(\xi)$ versus parameter $\nu$
at $A=10$. Black and red lines correspond to $B=20$ and $B=30$, respectively.
Panel~(a) represents the initial values $\chi_0$ determined from (\ref{ICs}).
Panel~(b) shows the values of first zero, when $\chi(\xi_B)=0$.
These are built at $\nu^{\mathrm{black}}_{\mathrm{min}}\simeq 2.066$ and
$\nu^{\mathrm{red}}_{\mathrm{min}}\simeq 1.805$ found numerically.}
\end{figure}

The possible dependencies of $\chi_0$ on $\nu$ are shown in
Fig.~\ref{chi0}~(a). We present also the values of $\xi_B$ in
Fig.~\ref{chi0}~(b), which are used in order to limit the system
size. Particular distributions of particles at $\xi\leq\xi_B$ can be
seen in Fig.~\ref{dens} (solid lines).

Fig.~\ref{chi0} reveals two different modes of model behavior, which
are separated by a turning point. We assume that the change of
regime is associated with a {\it phase transition}, which we will
study. Moreover, the system size in Fig.~\ref{chi0}~(b) does not
depend completely on parameters $A$ and $B$ of interactions (in
contrast to \cite{Harko2011}), but is affected also by $\nu$ which
defines the dominant phase of matter.

For a deeper understanding of the model properties, let us
define a chemical potential $\mu_q(\xi)$ as
\begin{equation}
\mu_q(\xi)+A\varphi(\xi)-\frac{1}{2\chi(\xi)}\Delta_\xi\chi(\xi)=u.
\end{equation}
Accordingly to the equation of motion (\ref{teq1}), $\mu_q$ determines
$\chi$ as
\begin{equation}\label{muQ}
\mu_q(\xi)=B\chi^4(\xi),\qquad \mu_q(\xi_B)=0.
\end{equation}

Replacing $\mu$ with $\mu_q$ in (\ref{GDr})--(\ref{Eu}),
we reproduce the expressions (\ref{EPtf}) for the internal
energy density $\varepsilon(\xi)$ and the local pressure $p(\xi)$
at $\eta=\chi^2$.

Internal pressure $p(\xi)$ evolves spatially as
\begin{equation}\label{qPev}
\frac{\partial_\xi p(\xi)}{\eta(\xi)}=\partial_\xi\mu_q(\xi)=-A\frac{n(\xi)}{\xi^2}
+\partial_\xi\left(\frac{1}{2\chi(\xi)}\Delta_\xi\chi(\xi)\right),
\end{equation}
what is in contrast with (\ref{Pev}) in the TF approximation. This
is a complicated hydrostatic equation (similar to Eq.~(15) in
\cite{Harko2019}), and we do not find its solution. Moreover, all
necessary fields can be found directly on the base of
(\ref{g2})--(\ref{g4}).

Nevertheless, (\ref{qPev}) says that internal pressure $p$ is
balanced by the resulting pressure caused by gravity and quantum
fluctuations. It is interesting to evaluate the contributions of
these effects.

To do this, we turn to more simple and known relations:
\begin{equation}
p=\frac{2}{3}B\chi^6=\frac{2}{3}\left[\frac{1}{2}\chi\Delta_\xi\chi+u\chi^2-A\chi^2\varphi\right],
\end{equation}
where (\ref{teq1}) is inserted.

Since the chemical potential $u$ consists here of two parts, accordingly to (\ref{g4}),
we introduce the following components of pressure:
\begin{eqnarray}
p_{\nu}&=&\frac{1}{3}\left(\chi\Delta_\xi\chi+\nu\chi^2\right),\\
p_{\mathrm{gr}}&=&-\frac{2}{3}A\eta\,\left[\upsilon(\xi_B)+\varphi(\xi)\right],
\end{eqnarray}
which give us $p_{\nu}+p_{\mathrm{gr}}=p$ by construction. We can see that
$p_{\nu}\sim\chi^2$ and corresponds to the wave fluctuations, regulated by
free parameter $\nu$, while $p_{\mathrm{gr}}\sim\chi^4$ and describes
pair interaction which is controlled by gravitational constant $G$ absorbed
by parameter $A$.

Average value of pressure $p_{gr}$ is given by expression:
\begin{equation}
VP_{\mathrm{gr}}=-\frac{2}{3}A\int_0^{\xi_B}\upsilon(\xi)[\upsilon(\xi_B)-\upsilon(\xi)]\,\rmd\xi<0.
\end{equation}
This means that the pressure created by interactions is $P-P_{\mathrm{gr}}>0$. At the
same time, $P_\nu>0$ supports the pressure inside the system. From another point of
view, action of $P_\nu$ (mean value of $p_\nu$) is equivalent to external influence
on the system.

Let us show how $p_\nu$ can be related with an external field $h$ of Landau theory~\cite{LL}.
First of all, we assume that the unperturbed theory is given by $\Gamma_{\mathrm{TF}}$.
Associating a perturbation with including quantum fluctuations and field $\nu$, given by hands,
a perturbation part of a total functional $\Gamma$ is presented in the form:
\begin{eqnarray}
2\frac{\Gamma_{\mathrm{pert}}}{\Gamma_0}&=&-\int_0^{\xi_B}\eta(\xi)h(\xi)\,\xi^2\rmd\xi,\\
h(\xi)&=&\frac{1}{\chi(\xi)}\Delta_\xi\chi(\xi)+\nu.
\end{eqnarray}
Therefore, the equation of motion here reads $(\delta\Gamma/\delta\eta(\xi))_{h=\mathrm{const}}=0$,
what leads to (\ref{teq1}). It immediately means also that the particle density $\eta(\xi)$
is an order parameter (conjugated to $h$) of such a theory. To assign a physical meaning to
this perturbation, we see a product $\eta(\xi)h(\xi)$ is nothing but the pressure $p_\nu$.

Often phase transitions are associated with changes of intensive parameters like temperature
and pressure (in the absence of magnetic and electric fields). Since $T=0$ in our model,
changes are stimulated by pressure. Microscopically, our assumption of a phase transition
in the model can now be justified by the presence of a perturbation $P_\nu$ related to
quantum fluctuations as well as the form of interaction that allows the existence of
two nonequivalent phases of matter.

\begin{figure}[htbp]
\begin{picture}(80,35)
\put(-70,-130){\includegraphics[width=8cm,angle=0]{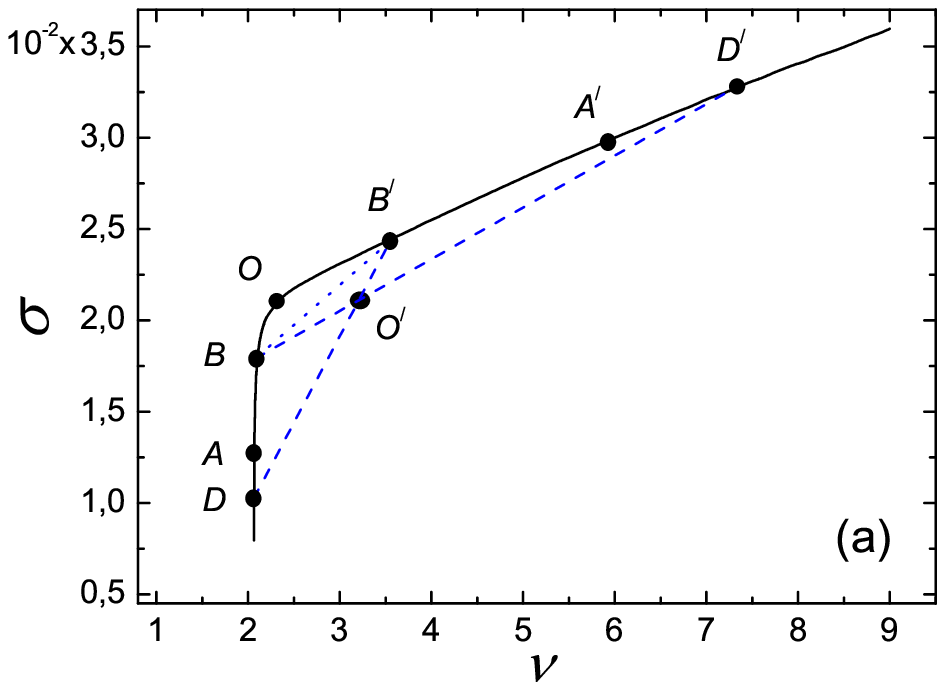}}
\put(-70,-284){\includegraphics[width=8.3cm,angle=0]{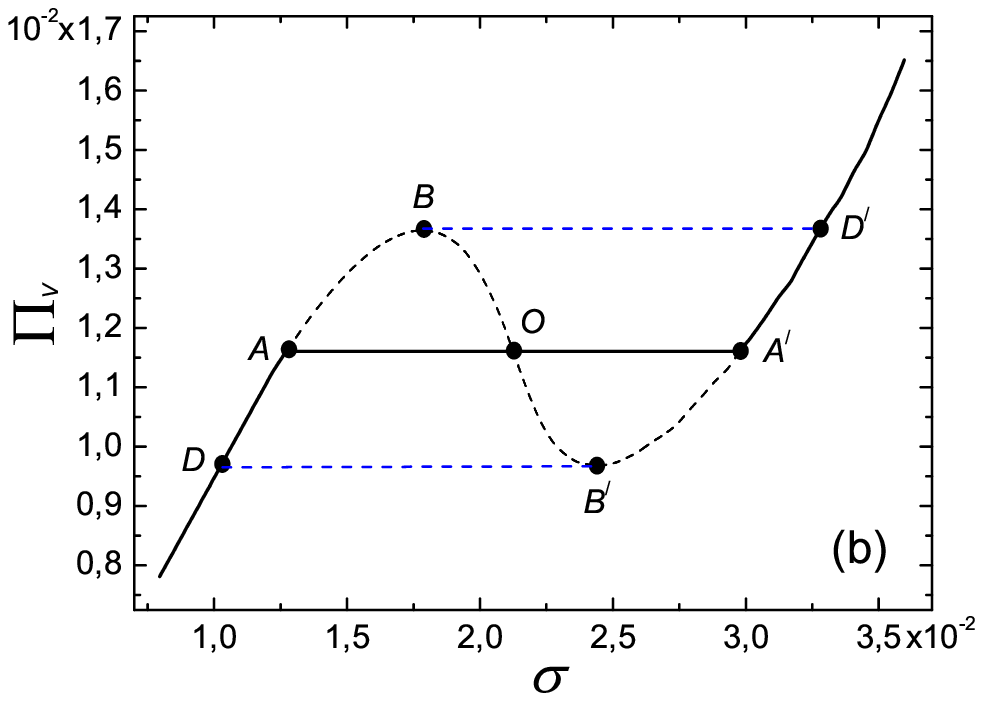}}
\end{picture}
\vspace*{93mm}
\caption{\label{sig} Discontinuity of mean density $\sigma$ at fixed $A=10$,
$B=20$ and at varying $\nu$ (a) and $\Pi_\nu$ (b). Panel~(b) shows
the metastable ($AB$ and $A^\prime B^\prime$) and unstable ($BB^\prime$)
states. Straight line $AA^\prime$ is constructed by the Maxwell rule.
The connections between key points in Panel~(b) are
{\it schematically} transferred to Panel~(a).}
%% $\sigma_A\simeq0.0127$, $\sigma_O\simeq0.0213$, $\sigma_{A^\prime}\simeq0.0297$
\end{figure}

\section{Thermodynamical Quantities and Two Phases of Dark Matter}

To analyze the macroscopic properties of the model, we use a set of dimensionless
quantities:
\begin{eqnarray}
P&=&\frac{2B}{\xi^3_B}\int_0^{\xi_B}\chi^6(\xi)\,\xi^2\rmd\xi,\\
\Pi_{\nu}&=&-\frac{(B/A)}{\xi^3_B}\int_0^{\xi_B}\left[(\partial_\xi\chi(\xi))^2-\nu\chi^2(\xi)\right]\xi^2\rmd\xi,\\
\sigma&=&\frac{3}{\xi^3_B}\int_0^{\xi_B}\chi^2(\xi)\,\xi^2\rmd\xi,\\
\tau&=&\frac{3}{\xi^3_B}\int_0^{\xi_B}(\partial_\xi\chi(\xi))^2\xi^2\rmd\xi,
\end{eqnarray}
which are obtained by averaging over volume $V=\xi^3_B/3$ of $p(\xi)$ (internal
pressure), $(B/A)\,p_{\nu}(\xi)$ (perturbation pressure rescaled for convenience),
$\eta(\xi)=\chi^2(\xi)$ (particle density), and $-\chi(\xi)\Delta_{\xi}\chi(\xi)$
(measure of fluctuations), respectively. Since the functions of our interest are
depended on $\nu$, $\chi_0$ and $\xi_B$, we also pay attention to their behavior.
We omit consideration of the quantities associated with the already given, such
as internal energy $E=VP/2$.

To convert the dimensionless quantities $\tau$, $\varepsilon$, $p$,
$P$ and $\Pi_\nu$ into physical units, we should multiply these by $\varepsilon_0$
from (\ref{eA}). The value of mean mass density is $\rho_0\sigma$.

Before a detailed description of the processes in the system, we want to check
the hypothesis of a phase transition. Since the only characteristic of matter
here is its density $\sigma$, let us trace its changes under different (external)
factors like $\nu$ and $\Pi_\nu$. Results of numerical
calculations are presented in Fig.~\ref{sig}. Note immediately that the
dependence of $\sigma$ on measure of fluctuations $\tau$ (given by graphic
$\sigma(\tau)=\{(\tau(\nu),\sigma(\nu))|\nu\geq\nu_{\mathrm{min}}\}$)
looks similar to Fig.~\ref{sig}~(a) and, thus, is omitted here.

First of all, we note a steep change of $\sigma(\nu)$ in
Fig.~\ref{sig}~(a) at $\nu_{\mathrm{min}}\simeq 2.066$. As was
argued above, $\chi(\xi)=0$ and, as a result, $\sigma=0$ at
$\nu<\nu_{\mathrm{min}}$, while $\sigma(\nu_{\mathrm{min}})>0$. It
can be interpreted as a first-order phase transition. This
phenomenon appears due to the rule of finding initial condition
$\chi_0$ (\ref{ICs}). However, this is not the subject of our study.
More interesting for us is a behavior of the matter near a turning
point $O$, that is, at $\nu\to\nu_{\mathrm{min}}$.

To understand the processes near the point $O$, we appeal to
Fig.~\ref{sig}~(b). Indeed, backbending in Fig.~\ref{sig}~(b)
reveals the first-order phase transition, which exhibits existence
of ``gaseous'' and ``liquidlike'' phases of the matter. This process
differs from the familiar boiling water due to our consideration of
the quantum system at $T=0$ and the mechanism which is based on
quantum fluctuations or, alternatively, associated with compression
$\Pi_\nu$ (that may be convenient further when considering
variations in macroscopic parameters).

\begin{figure}[h]
\begin{tabular}{cc}
\includegraphics[width=0.485\linewidth]{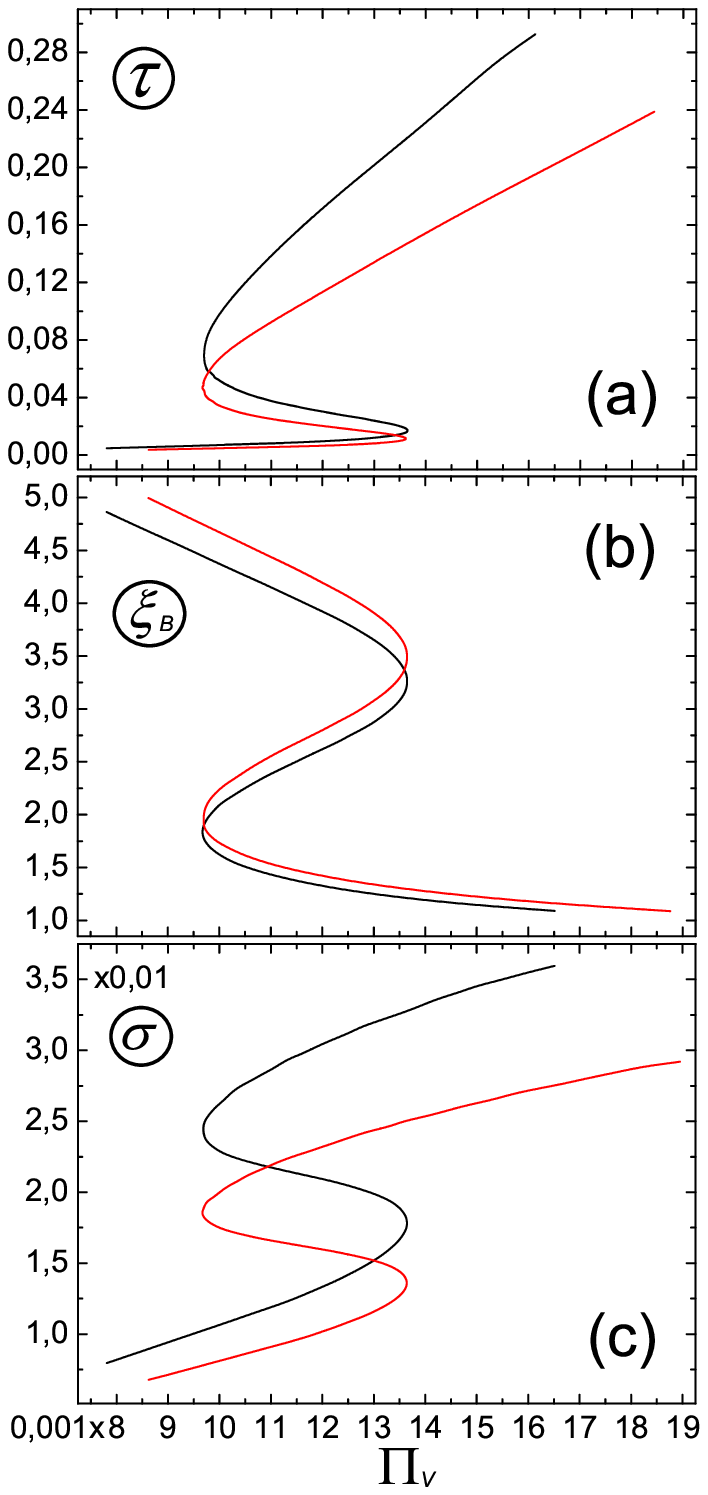}&
\includegraphics[width=0.487\linewidth]{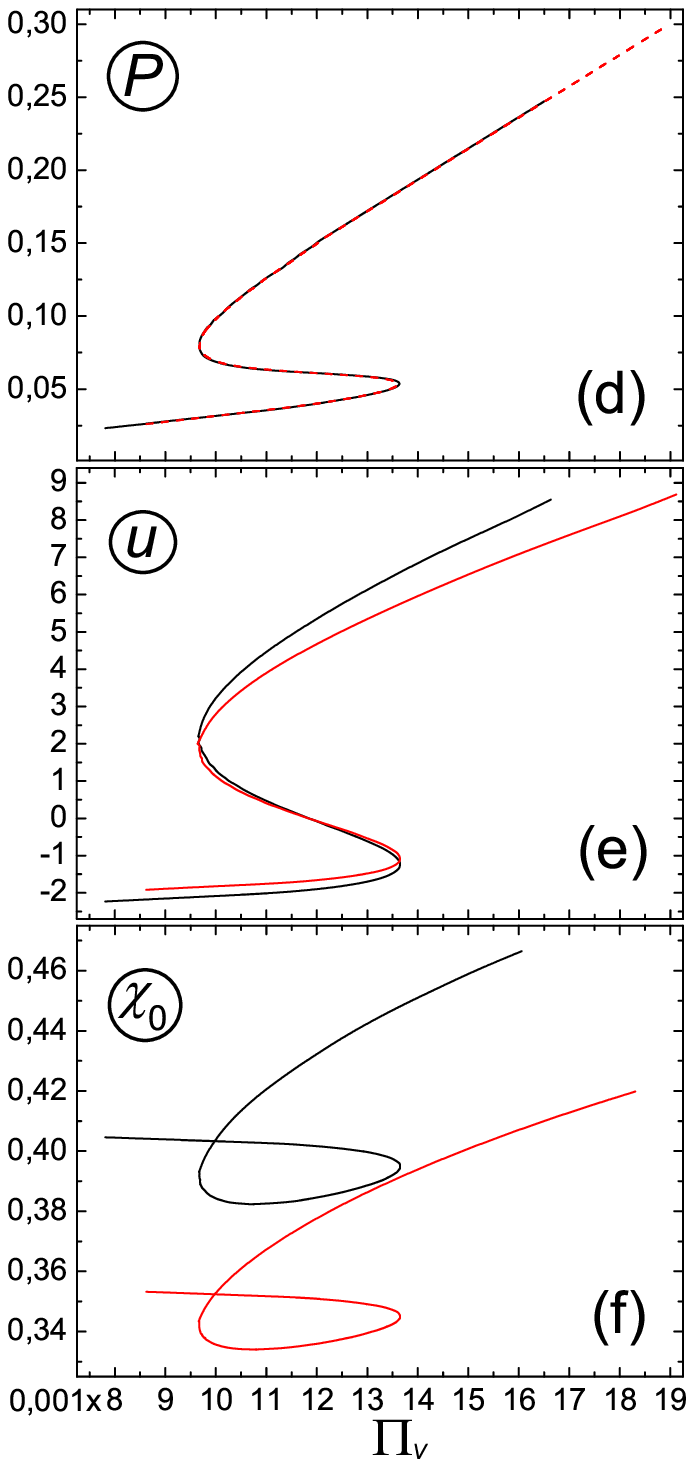}
\end{tabular}
\vspace*{-2mm}
\caption{\label{funcs} Dependence of dimensionless characteristics on $\Pi_\nu$
at $A=10$. Black and red lines correspond to $B=20$ and $B=30$, respectively.}
\end{figure}

The characteristics, computed numerically, are depicted in
Fig.~\ref{funcs} and behave as multivalued functions of $\Pi_\nu$.
These graphics represent isotherms at $T=0$. Varying the parameter
$B$ (and $A$), we can conclude that the region of backbending
presence, projected onto $\Pi_\nu$ axis, remains the same due to
scale factor $B/A$, appeared by defining $\Pi_\nu$. It means that
there is no possibility to achieve the critical point of phase
transition, when $\partial\Pi_\nu/\partial f=0$ and
$\partial^2\Pi_\nu/\partial f^2=0$ simultaneously for any $f$ of
these functions, by changing $B$ (and $A$). Comparing with water, in
this model there is no natural parameter analogous to temperature,
the increase of which would lead to boiling. We admit that by
considering $T>0$, one cannot also observe a similar process,
because a vaporization temperature here is expected to be much
higher than the critical temperature of the condensate and to be
comparable (in the energy units) with the gravitational energy of
the system. Nevertheless, Fig.~\ref{funcs} allows us to extract the
conditions of transition between gaseous and liquid-like phases at
$T=0$, if such a situation is realized in nature.

\begin{figure}[htbp]
\includegraphics[width=7.0cm,angle=0]{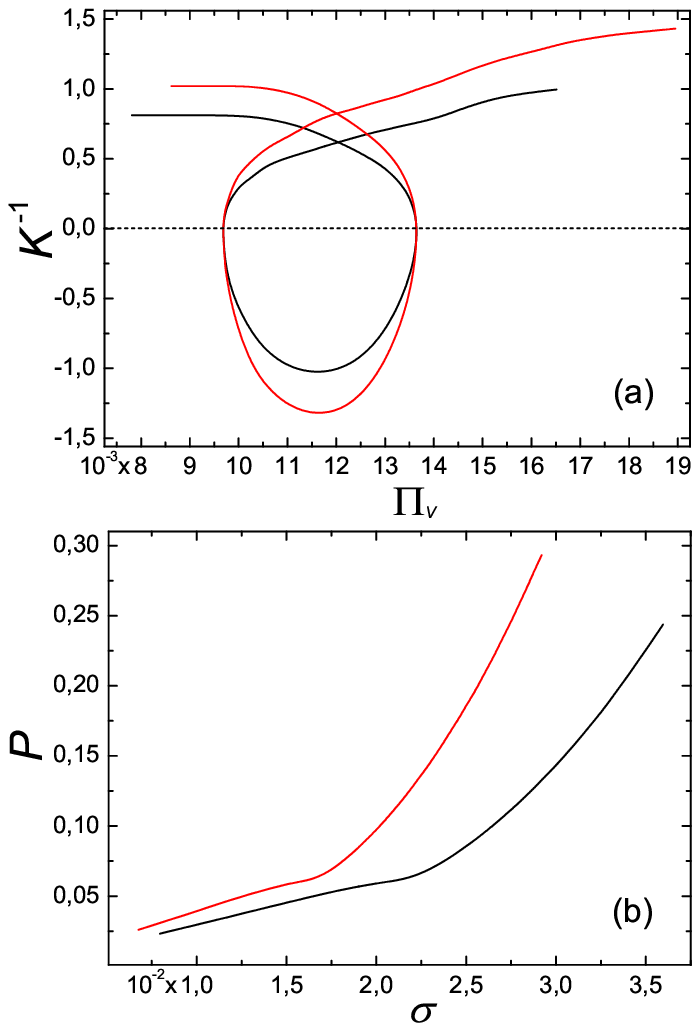}
\vspace*{-2mm}
\caption{\label{XscP} Behavior of incompressibility $K^{-1}$
and internal pressure $P$ at $A=10$. Black and red lines correspond
to $B=20$ and $B=30$, respectively.}
\end{figure}

Considering $\Pi_\nu$ as external compression for convenience and
omitting a physical interpretation of the region of instability
(like $BB^\prime$ in Fig.~\ref{sig}), Fig.~\ref{funcs} permits to
conclude the following. By intensifying the compression $\Pi_\nu$,
starting from the minimum, we can see the size $\xi_B$ (b) of the
system decreases, and the mean density $\sigma$ (c) increases
throughout the process. At the same time, the central density
$\chi^2_0$ (see (f)) and fluctuations $\tau$ (a) do not noticeably
change at the first stage, but begin to grow significantly after the
transition to a more dense, liquidlike phase. A similar behavior is
observed for chemical potential $u$ (e) and internal pressure $P$
(see (d), where curves with different $B$ overlap). As it could be
expected the point of two phases coexistence, $u=0$, lies in the
instability region.

Note that in all the figures wherein the dependence on compression
$\Pi_\nu$ is involved, we observe the common feature
--- existence of the same pair of special/distinguished values of
$\Pi_\nu$, that is $\Pi_\nu^{(1)}\simeq 9.65\cdot 10^{-3}$ and
$\Pi_\nu^{(2)}\simeq 13.65\cdot 10^{-3}$ in dimensionless units
(or, in physical units,  $\Pi_\nu^{(1)}\simeq 5.23\cdot
10^{-14}\ \mathrm{Pa}$ and $\Pi_\nu^{(2)}\simeq 7.40\cdot 10^{-14}\ \mathrm{Pa}$).
Due to that, the curves are separated in three regions of qualitatively
differing behavior: two ``normal'' parts filled with stable states of
the system, which correspond to compressions such that
 $\Pi_\nu \leq \Pi_\nu^{(1)}$ and $\Pi_\nu \geq \Pi_\nu^{(2)}$
respectively, and the interval/region of instability. As seen from
the figures, for the latter interval there appears an obvious
backbending which shows the opposite behavior of the considered
functions as compared to the normal (stable) parts of the curves.

Our results predict significant fluctuations $\tau$ in the denser
state of dark matter. Physically, this is due to the increasing role
of repulsion $\chi^6$ among particles by growing $\sigma$. Although,
associating $\tau$ with the effective temperature, the liquidlike
phase of DM can be assumed to be in thermalized state what is
already noted in \cite{Kamada} within the SIDM model. A simultaneous
increase in ``temperature'' $\tau$ and density $\sigma$ resembles
the behavior of water in the temperature range $t=0\div4{}^\circ C$.

Isothermal compressibility $K$, playing the role of susceptibility,
and its inverse $K^{-1}$ defined here as
\begin{equation}
K=\frac{\partial\sigma}{\partial\Pi_\nu},\qquad
K^{-1}=\frac{\partial\Pi_\nu}{\partial\sigma},
\end{equation}
are also important characteristics and can be computed numerically
(with high precision loss) by using dependencies in
Fig.~\ref{funcs}~(c). The result is sketched in Fig.~\ref{XscP}~(a).

%%%%%%%%%%%%%%%%%%%%%%%%%%%%%%%%%%%%%%%%%

%A negative part of incompressibility, $K^{-1}<0$, corresponds to the
%instability region and, therefore, can be neglected in the physical
%picture. On the other hand, a positive part, $K^{-1}>0$, consists of
%two branches which describe the phases of matter with their
%individual properties.

{The most important positive (or regular) part, with $K^{-1}>0$,
consists of two branches which describe the two distinct phases of
matter with their individual properties. It is interesting to remark
that the incompressibility behaves like a constant for lower values
of compression, say for $\Pi_\nu \leq \Pi_\nu^{(1)}$, whereas it
shows rising property for higher values when $\Pi_\nu$ grows,
including the region with $\Pi_\nu > \Pi_\nu^{(2)}$.}

{On the other hand, there is a negative part of incompressibility,
$K^{-1}<0$, that means the instability region, which corresponds
to the intermediate values of compression, i.e.
for $\Pi_\nu^{(1)} < \Pi_\nu < \Pi_\nu^{(2)}$.
Using equilibrium statistics, this is usually neglected in the physical
picture.  However, as shown in some papers, there exist in reality
the structures (materials) characterized by negative $K$, see e.g.  %systems
\cite{neg.sss-1,neg.sss-2,neg.sss-3,neg.sss-4}. Besides, as some
works demonstrate, the vanishing value of the effective speed of
sound (that would happen in our model at $\Pi_\nu = \Pi_\nu^{(1)}$
and $\Pi_\nu = \Pi_\nu^{(2)}$), may be also of importance
\cite{zero.ss1,zero.ss2}. Therefore, the peculiar instability
region, together with vanishing property in some points, may be
taken into account more seriously in the context of the present
model, and the physical meaning of all these unusual properties
deserves more detailed study.}

%%%%%%%%%%%%%%%%%%%%%%%%%%%%%%%%%%%%%%%%%

Dependence of internal pressure $P$ (and internal energy $E$) on
mean density $\sigma$ in Fig.~\ref{XscP}~(b) indicates rather
a second-order phase transition with a turning point at
$\partial^2P/\partial\sigma^2=0$. Studying numerically the
dependencies $P$ on $\sigma$ for different values of $A$ and $B$,
which are not shown here, the presence of a backbending (accompanying
a first-order phase transition) was not confirmed.

Thus, the visualized equation of state also reveals two phases.
However, it does not permit us to explain in details a mechanism of
phase transition. Moreover, increasing a parameter $B$, the system
jumps into liquid-like phase at smaller values of $\sigma$.
 This probably witnesses enhanced role of three-particle interaction.

{ There is one further feature of the role of parameter
$B$ (at fixed $A<B$) clearly seen from Figs.~2,3,5,6.
 Namely, its influence on the studied thermodynamical quantities is the
following: the greater is (the value of) $B$ the higher are
%(at fixed other parameters)
the quantities $\xi_B(\Pi_\nu)$, $K^{-1}(\Pi_\nu)$ and $P(\sigma)$;
on the contrary, regularity is inverse for the density $\eta(\xi)$,
$\chi_0(\Pi_\nu)$, $\chi_0(\nu)$, $\xi_B(\nu)$, $\tau(\Pi_\nu)$,
$u(\Pi_\nu)$ and $\sigma(\Pi_\nu)$.
 At last, there is a special case --- the quantity $P(\Pi_\nu)$, see Fig.~5d,
 which shows (almost) no dependence on the value of $B$.}

Adiabatic speed of sound $c_s$ can be found (in physical units) as
\begin{equation}
c_s=\left.\sqrt{\frac{\varepsilon_0}{\rho_0}\frac{\partial
p}{\partial\eta}}\,\right|_{\eta=\chi^2_0} =\frac{\hbar\sqrt{2B}}{m
r_0}\,\chi^2_0.
\end{equation}
Thus, the magnitude of $c_s$ is evaluated as
\begin{eqnarray}
c_s&\simeq& 2.36\cdot10^6\ \mathrm{cm}\,\mathrm{s}^{-1}\,\left[\frac{A}{10}\right]^{-1/4}
\,\left[\frac{B}{20}\right]^{1/2}\,\left[\frac{\chi_0}{0.4}\right]^2
\nonumber\\
&&
\times \left[\frac{mc^2}{10^{-22}\ \mathrm{eV}}\right]^{-1/2}
\,\left[\frac{\rho_0}{10^{-20}\ \mathrm{kg}\,\mathrm{m}^{-3}}\right]^{1/4}.
\label{SS}
\end{eqnarray}
This value is in accordance with the predictions of other models~\cite{Harko2019}.

%%%%%%%%%%%%%%%%%%%%%%%%%%%%%%%%%%%%%%%%%%%%%%%%%%%%%%%%%%%%%%%%%%%%%
\section{Concluding remarks}

%\vspace{1mm}

The results obtained above witness clearly that the inclusion of
sixth order repulsive self-interaction of ultralight dark matter
within the particular modification of Gross--Pitaevskii equation
%(with nonlocal way of account of gravity),
leads to highly nontrivial properties of BEC DM and thus to unexpected
galactic core structure and dynamics.
The main results of the study, obtained numerically, are presented in
the form of Figs.~3--6, equipped with comments in the text, which we
do not repeat here.

%\vspace{1mm}

There is a principal issue of whether the BEC dark bosons are
 viewed as {\it elementary or composite}: the answer to this could
 give some guess towards revealing the nature of dark matter.
  If deformed bosons are dealt with, then (i) some additional interaction between
 bosons, besides the familiar pure bosonic attraction of quantum-statistical origin,
 can be  effectively taken into  account~\cite{Scarfone},
 also (ii) a rather simple effective account of compositeness
 aspects is as well at our disposal~\cite{GKM1,GKM2,GM1}, and (iii)
 these two issues can be treated jointly~\cite{GM2,GM3,GM4}.
It is important to emphasize that compositeness and the related
deformation significantly affect~\cite{muBose1,Kurzyn} the critical
temperature of condensation, depending on the constituent particles
so that the case of Bose-Bose composites basically differs from that
of Fermi-Fermi composites.

As mentioned in the Introduction, the nontrivial phase structure of
the core in the central part of DM halo was noticed in the work of
Chavanis~\cite{Chavanis2} (namely the existence of dilute and dense
phases, and the related {\it zero's order} phase transition seen
through the mass-radius relation).
Unlike that, in the present work we have demonstrated both the presence
of two phases (consisting of stable, metastable, and unstable branches)
and of the {\it first-order} phase transition.

It was already pointed out that phase transitions in dark matter can
trigger {\it gravitational waves}~\cite{Schwaller,Croon,Bezares},
see also~\cite{Bramante}. Especially this can hold if one takes into
account that the nontrivial phase structure can even more naturally
occur when we deal with the colliding (or merger of) two galaxies,
see e.g.~\cite{Bezares,Lee-Lim-Choi,Maleki}. In this context, it
would be interesting to investigate possible generation of
gravitational waves caused by the phase transitions (in DM core) of
the form described in this paper, possibly with extension of the
model through inclusion of dynamical aspects. Till that will be
performed, the results described above, in particular those
visualized in Figs. 3-6, can be viewed as a ``screen-shot'' (or
fixed-time slice) of the whole picture. The complete dynamics
(continuous time evolution) is obviously needed, and separate work
will be devoted to this subject.

No doubt of importance is the necessity to consistently study the
simultaneous presence of both quartic and sextic terms in the scalar
potential. If both these terms are repulsive, the inclusion of the
4th order term should somewhat enhance the effects of the sixth
order self-interaction term studied above. Not less interesting and
important task is to establish the exact nature (structure) of the
appearing two distinct phases manifested by the DM core. To this
end, the role of three-particle interactions should be studied
within the quantum-mechanical framework. In this context, knowledge
of the second and third virial coefficients of
$\tilde\mu,q$-deformed Bose gas model~\cite{GM4} could be helpful.
We hope to report on such results elsewhere.

\vspace{2mm}

\section*{Acknowledgements}

A.M.G. acknowledges support from the National Academy of Sciences of
Ukraine by its priority project No. 0120U100935 "Fundamental
properties of the matter in the relativistic collisions of nuclei
and in the early Universe". The work of A.V.N. was supported by the
project No. 0117U000238 of the National Academy of Sciences of
Ukraine.

%% %% %% %% %% %% %% %% %% %% %% %% %% %% %% %% %% %% %% %% %% %% %% %% %% %% %% %%


\begin{thebibliography}{99}

\bibitem{Zwicky}
F.~Zwicky, Astrophys. J. {\bf 86}, 217 (1937). % doi:10.1086/143864.

\bibitem{Bertone}
G.~Bertone and D.~Hooper, Rev. Mod. Phys. {\bf 90}, 045002 (2018).
% doi:10.1103/RevModPhys.90.045002   %rev-1

\bibitem{Sin}
S.-J.~Sin, Phys. Rev. D {\bf 50}, 3650 (1994).
 %%https://doi.org/10.1103/PhysRevD.50.3650

\bibitem{Lee}
J.-W.~Lee and I.-G.~Koh, Phys. Rev. D {\bf 53}, 2236 (1996).
   %%https://doi.org/10.1103/PhysRevD.53.2236

\bibitem{Hu}
W.~Hu, R.~Barkana, and A.~Gruzinov, Phys. Rev. Lett. {\bf 85}, 1158 (2000).
 %https://doi.org/10.1103/PhysRevLett.85.1158

\bibitem{Matos}
A.~Suarez, V.~Roblez, and T. Matos, Astrophys. Space Sci.
Proc. {\bf 38}, 107 (2014).

\bibitem{Lee17}                                           %rev-3
J.-W.~Lee, EPJ Web Conf. 168, 06005 (2018).

\bibitem{Hui}                %rev-4
L.~Hui, J.P.~Ostriker, S.~Tremaine, and E.~Witten, Phys. Rev. D {\bf 95}, 043541 (2017).
% arXiv: 1610.08297.

\bibitem{Urena}               %rev-5
L.A.~Urena-Lopez, EPJ Web Conf. 168, 06005 (2018).

\bibitem{Ferreira}               %rev-6
E.G.M.~Ferreira, {\it Ultra-Light Dark Matter}, astro-ph/2005.03254.

\bibitem{Fan}                     %rev-7
J.J.~Fan, Phys. Dark Univ.  {\bf 14},  84 (2016).

\bibitem{Peebles}
P.J.E.~Peebles,  Astrophys. J. Lett. {\bf 534}, L127 (2000).

\bibitem{Berezhiani}
L.~Berezhiani and J.~Khoury, Phys. Rev. D {\bf 92}, 103510 (2015).

\bibitem{Goodman}
J.~Goodman, New Astron. {\bf 5}, 103 (2000).

\bibitem{Bohmer}
C.G.~Bohmer and T.~Harko, J. Cosmol. Astropart. Phys. {\bf 06}, 025 (2007).

\bibitem{Chavanis1}  %%logotropic
P.H.~Chavanis, Eur. Phys. J. Plus {\bf 132}, 248 (2017).

\bibitem{Chavanis2}       %% cos
P.H.~Chavanis, Phys. Rev. D {\bf 98}, 023009 (2018).

\bibitem{Sahni}            %% cosh
V.~Sahni and L.~Wang, Phys. Rev. D {\bf 62}, 103517 (2000).
% astro-ph/9910097.

\bibitem{Matos2}
T.~Matos and L.A.~Urena-Lopez, Class. Quant. Grav. {\bf 17}, L75 (2001).

\bibitem{Guzman}
M.~Alcubierre, F.S.~Guzman, T.~Matos, D.~Nunez, L.A.~Urena-Lopez, and P.~Wiederhold,
Class. Quant. Grav. {\bf 19}, 5017 (2002).

\bibitem{CDM}
G.R.~Blumenthal, S.M.~Faber, J.R.~Primack, and M.J.~Rees,
Nature {\bf 311}, 517 (1984).
% doi:10.1038/311517a0.

\bibitem{Harko2011}
T.~Harko, J. Cosmol. Astropart. Phys. {\bf 05}, 022 (2011).

\bibitem{Deng}
H.~Deng, M.P.~Hertzberg, M.H.~Namjoo, and A. Masoumi,
Phys. Rev. D {\bf 98}, 023513 (2018).
% arXiv: 1804.05921v2.

\bibitem{Khlopov}
M.Yu.~Khlopov, B.A.~Malomed and Ya.B.~Zeldovich,
 %Gravitational instability of scalar field and primordial black holes.
Mon. Not. R. Astron. Soc. {\bf 215}, 575, (1985). %%-589.

\bibitem{Harko2}
T.~Harko, Phys. Rev. D {\bf 89}, 084040 (2014).

\bibitem{Salucci}
P.~Salucci, %% The distribution of dark matter in galaxies
 Astron. Astrophys. Rev. {\bf 27}, 2 (2019).

\bibitem{Diez2014}
A.~Diez-Tejedor, A.X.~Gonzalez-Morales, and S.~Profumo,
Phys. Rev. D {\bf 90}, 043517 (2014).

\bibitem{Zang2018}
X.~Zhang {\it et al.}, Eur. Phys. J. C {\bf 78}, 346 (2018).

\bibitem{Kun2020}
E.~Kun, Z.~Keresztes, and L.~Gergely, Astron. Astrophys. {\bf 633}, A75 (2020).

\bibitem{Craciun2019}
M.~Craciun and T.~Harko, Roman. Astron. J. {\bf 29}, 109 (2019).

\bibitem{Castellanos2020}
E.~Castellanos, C.~Escamilla-Rivera, Int. J. Mod. Phys. D 29,
2050063 (2020).

\bibitem{BBBS}
N.~Bar, D.~Blas, K.~Blum, and S.~Sibiryakov, Phys. Rev. D {\bf 98}, 083027 (2018).
% arXiv: 1805.00122.

\bibitem{Lee2016}
J.-W.~Lee, Phys. Lett. B {\bf 756}, 166 (2016).

\bibitem{Schive1}
H.-Y.~Schive, M.-H.~Liao, T.-P.~Woo, S.-K.~Wong, T.-H.~Chiueh, T.~Broadhurst,
and W-Y.~Pauchy Hwang, Phys. Rev. Lett. {\bf 113}, 261302 (2014).
% arXiv: 1407.7762.

%%%%%%%%%%%%%%%%
\bibitem{axion}
P.~Sikivie and Q.~Yang, Phys. Rev. Lett. {\bf 103}, 111301 (2009).

\bibitem{axion2}
T.~Noumi, K.~Saikawa, R.~Sato, and M.~Yamaguchi,
Phys. Rev. D {\bf 89}, 065012 (2014).

\bibitem{axion3}
S.~Davidson, Astropart. Phys. {\bf 65}, 101 (2015).
%  arXiv: 1405.1139.
%%https://doi.org/10.1016/j.astropartphys.2014.12.007

\bibitem{axion4}
A.H.~Guth, M.P.~Hertzberg, and C.~Prescod-Weinstein,
Phys. Rev. D {\bf 92}, 103513 (2015).

\bibitem{axion5}
E.D.~Schiappacasse and M.P.~Hertzberg, JCAP {\bf 1801}, 037 (2018).
% arXiv: 1710.04729.

\bibitem{Stue}
T.R.~Govindarajan and N.~Kalyanapuram,
Mod. Phys. Lett. A  {\bf 34}, 1950330 (2019).
% arXiv: 1902.08768. %[hep-ph]

\bibitem{Das}
S.~Das and R.K.~Bhaduri, Class. Quant. Grav. {\bf 32}, 105003 (2015).
% arXiv: 1411.0753.
%https://doi.org/10.1088/0264-9381/32/10/105003

\bibitem{Kun}
E.~Kun, Z.~Keresztes, S.~Das, and L.A.~Gergely,
Symmetry {\bf 10}, 520 (2018).

\bibitem{deRham}
C. de Rham, J.T.~Deskins, A.J.~Tolley, and S.-Y.~Zhou,
Rev. Mod. Phys. {\bf 89}, 025004 (2017).
%https://doi.org/10.1103/RevModPhys.89.025004

\bibitem{Mirza}
Z.~Ebadi, B.~Mirza, and H. Mohammadzadeh, JCAP {\bf 11}, 057 (2013).
 %%https://doi.org/10.1088/1475-7516/2013/11/057

\bibitem{muBose1}
A.M.~Gavrilik, I.I.~Kachurik, M.V.~Khelashvili, and A.V.~Nazarenko,
Physica A: Stat. Mech. Applic. {\bf 506}, 835 (2018).

\bibitem{muBose2}
A.M.~Gavrilik, I.I.~Kachurik, and M.V.~Khelashvili,
Ukr. J. Phys. {\bf 64}, 1042 (2019).
 %%DOI: https://doi.org/10.15407/ujpe64.11.1042

\bibitem{Nazar}
A.V.~Nazarenko, Int. J. Mod. Phys. D {\bf 29}, 2050018 (2020).
 %%https://doi.org/10.1142/S0218271820500182

%%%%%%%%%%%%%%%%%%%
\bibitem{Luckins}
E.K.~Luckins and R.A. Van Gorder, Ann. Phys. {\bf 388}, 206 (2018). %206-234
 %%https://doi.org/10.1016/j.aop.2017.11.009

\bibitem{Chen2017}
S.-R.~Chen, H.-Yu~Schive, and T.~Chiueh, MNRAS {\bf 468}, 1338
(2017).
% arXiv: 1606.09030v2.

\bibitem{Schive2014}
H.-Yu~Schive, T.~Chiueh, and T.~Broadhurst, Nat. Phys. {\bf 10}, 496
(2014).
% arXiv: 1406.6586v1.

%%%[8] T. Harko, JCAP 05 (2011) 022.

\bibitem{Horedt}
G.P.~Horedt, Astron. Astrophys. {\bf 160}, 148 (1986).

\bibitem{LL}
L.D.~Landau and E.M.~Lifshitz, {\it Statistical Physics} (Pergamon
Press, NY, 1978).

\bibitem{Harko2019}
T.~Harko, Eur. Phys. J. C {\bf 79}, 787 (2019).
% arXiv: 1909.05022.

\bibitem{Kamada}      %[11]
A.~Kamada, M.~Kaplinghat, A.B.~Pace, and H.-B.~Yu,
Phys. Rev. Lett. {\bf 119}, 111102 (2017).
% arXiv: 1611.02716.
%%https://doi.org/10.1103/PhysRevLett.119.111102

\bibitem{neg.sss-1} %negative speed of sound squared - 1
R.~Lakes and K.W.~Wojciechowski, Phys. Status Solidi {\bf 245}, 545
(2008).
 %https://doi.org/10.1002/pssb.200777708

\bibitem{neg.sss-2} %%negative speed of sound squared - 2
R.~Gatt and J.N.~Grima, Phys. Status Solidi, {\bf 2}, 236 (2008).
  %% https://doi.org/10.1002/pssr.200802101

\bibitem{neg.sss-3} %%negative speed of sound squared - 3
J.A.~Kornblatt, E.B.~Sirota, H.E.~King Jr., H.~Baughman, C.-X.~Cui, S.~Stafstrom,
and S.O.~Dantas, Science {\bf 281}, 143 (1998).
    %% DOI: 10.1126/science.281.5374.143a

\bibitem{neg.sss-4}  %%negative speed of sound squared - 4
B.~Moore, T.~Jaglinski, D.S.~Stone, and R.S.~Lakes, Philos. Mag.
Lett. {\bf 86}, 651 (2007).
   %% https://doi.org/10.1080/09500830600957340

\bibitem{zero.ss1} %zero speed of sound
C.-J.~Gao, M.~Kunz, A.R.~Liddle, and D.~Parkinson, Phys. Rev. D {\bf
81}, 043520 (2010).
  %% https://doi.org/10.1103/PhysRevD.81.043520

\bibitem{zero.ss2}
O.~Luongo and H.~Quevedo, Int. J. Mod. Phys. D {\bf 23}, 1450012
(2014).
 %% https://doi.org/10.1142/S0218271814500126

\bibitem{Scarfone}
A.M.~Scarfone and P. Narayana Swamy, J. Stat. Mech.: Theory \&
Experim. (2009) P02055.

\bibitem{GKM1}
A.M.~Gavrilik, I.I.~Kachurik, and Yu.A.~Mishchenko,
Ukr. J. Phys. {\bf 56}, 948 (2011).

\bibitem{GKM2}
A.M.~Gavrilik, I.I.~Kachurik, and Yu.A.~Mishchenko,
J. Phys. A: Math. Theor. {\bf 44}, 475303 (2011).
  %%ttps://doi.org/10.1088/1751-8113/44/47/475303

\bibitem{GM1}
A.M.~Gavrilik and Yu.A.~Mishchenko, Phys. Lett. A {\bf 376}, 1596 (2012).
  %%https://doi.org/10.1016/j.physleta.2012.03.053

\bibitem{GM2}
A.M.~Gavrilik and Yu.A.~Mishchenko, Ukr. J. Phys. {\bf 58}, 1171 (2013).

\bibitem{GM3}
 A.M.~Gavrilik and Yu.A.~Mishchenko, Nucl. Phys. B {\bf 891}, 466 (2015).
  %https://doi.org/10.1016/j.nuclphysb.2014.12.

\bibitem{GM4}
A.M.~Gavrilik and Yu.A.~Mishchenko,  Phys. Rev. E {\bf 90}, 052147 (2014).
   %https://doi.org/10.1103/PhysRevE.90.052147

\bibitem{Kurzyn}
S.-Y.~Lee, J.~Thompson, S.~Raeisi, P.~Kurzynski, and D.~Kaszlikowski,
New J. Phys. {\bf 17} (2015).

\bibitem{Schwaller}
P.~Schwaller, Phys. Rev. Lett. {\bf 115}, 181101 (2015).

\bibitem{Croon}
D.~Croon, V.~Sanz, and G.~White,
   %Model discrimination in gravitational wave spectra from dark phase transitions,
 J. High Energy Phys. 08 (2018)203.  %%arXiv: 1806.02332.   https://doi.org/10.1007/JHEP08(2018)203.

\bibitem{Bezares}
M.~Bezares and C.~Palenzuela, Class. Quant. Grav. {\bf 35}, 23 (2018).
  %%https://doi.org/10.1088/1361-6382/aae87c

\bibitem{Bramante}
 A.~Bhoonah, J.~Bramante, S.~Nerval and N.-Q.~Song,
 {\it Gravitational Waves From Dark Sectors, Oscillating
Inflatons, and Mass Boosted Dark Matter},
 arXiv: 2008.12306.

\bibitem{Lee-Lim-Choi}
J.-W.~Lee, S.~Lim, and D.~Choi, {\it BEC dark matter can explain
collisions of galaxy clusters}, arXiv: 0805.3827.

\bibitem{Maleki}
A.~Maleki, S.~Baghram, and S.~Rahvar, Phys. Rev. D {\bf 101}, 023508 (2020).
% arXiv: 1911.00486.


\end{thebibliography}
\end{document}